# The CMS Tracker Readout Front End Driver

C. Foudas, R. Bainbridge, D. Ballard, I. Church, E. Corrin, J.A. Coughlan, C.P. Day,
E.J. Freeman, J. Fulcher, W.J.F. Gannon, G. Hall, R.N.J. Halsall, G. Iles, J. Jones,
J. Leaver, M. Noy, M. Pearson, M. Raymond, I. Reid, G. Rogers, J. Salisbury, S. Taghavi, I.R.
Tomalin, O. Zorba

*Abstract--* **The Front End Driver, FED, is a 9U 400mm VME64x card designed for reading out the Compact Muon Solenoid, CMS, silicon tracker signals transmitted by the APV25 analogue pipeline Application Specific Integrated Circuits. The FED receives the signals via 96 optical fibers at a total input rate of 3.4 GB/sec. The signals are digitized and processed by applying algorithms for pedestal and common mode noise subtraction. Algorithms that search for clusters of hits are used to further reduce the input rate. Only the cluster data along with trigger information of the event are transmitted to the CMS data acquisition system using the S-LINK64 protocol at a maximum rate of 400 MB/sec. All data processing algorithms on the FED are executed in large on-board Field Programmable Gate Arrays. Results on the design, performance, testing and quality control of the FED are presented and discussed.**

## I. THE CMS SILICON TRACKER READOUT

THE CMS experiment is a general purpose detector with hermetic coverage designed for the Large Hadron Collider at CERN. CMS has been optimized to search for the Higgs particle which is associated with the origin of mass. The CMS Silicon Tracker measures charged particles by their bending in a 4T magnetic field using finely segmented silicon microstrip detectors read out by low noise APV25 analogue pipeline Application Specific Integrated Circuits (ASICS) [1]. A schematic diagram of the CMS Silicon Tracker readout system is shown in Fig. 1. The signals produced when charged particles pass through the Silicon Tracker are processed and stored on the APV25 ASICs. Each APV25 serves 128 silicon strips. The data are kept in the APV25s analogue pipelines until the decision from the CMS Level-1 trigger system is received. On a positive decision, the data from multiplexed pairs of APV25s are transmitted serially at 40 MHz via 65 m analogue optical links [2] to the FED cards in the CMS counting room. The data frame from each fibre consists of a 24 word header with the pipeline addresses of the data plus two error bits followed by the 256 data words. The entire CMS silicon tracker consists of about 9 million strips which will be read-out by 430 FEDs.

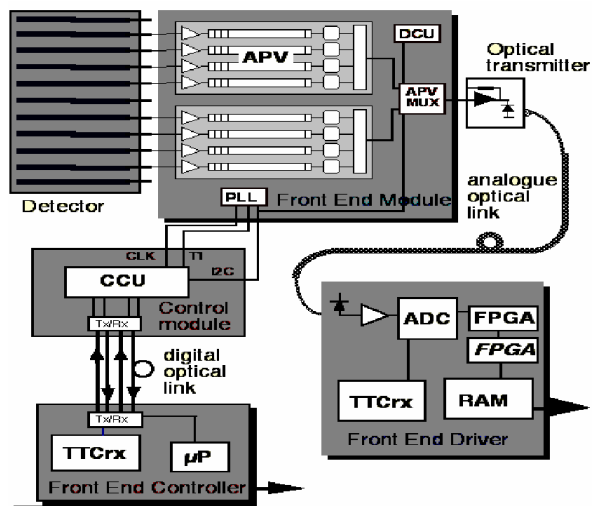

Fig. 1: The CMS Silicon Tracker Readout system. Charge depositions in the silicon strips (yellow) are read out by the APV25 ASCIS located on the detector. The data from pairs of APV25s are multiplexed and sent via analogue optical links to the FEDs (blue) located outside of the detector. The CCU and TTC system (red) delivers the trigger and clock. From the TTC system the Trigger is also sent to the FEDs.

## II. THE FED

### A. The FED Front-End

The main function of the FED, shown in Fig. 2, is to reduce the data produced by the silicon tracker by removing data that does not correspond to particle tracks such as pedestals and noise in general. Each FED receives data from 96 fibres each transmitting data from two APV25s. The FED front-end consists of eight identical units each one serving 12 fibres. Each front-end unit converts the data from optical to copper using one Optical Receiver module (Opto-Rx) [3]. The data are digitized using 6 dual 10 bit AD9218 ADCs [4] mounted on both sides of the card and clocked at 40 MHz. The gain and DC-offset of the Opto-Rx output signals are VME programmable in order to match the ADC ranges. The phases of the ADC clocks are finely adjusted with programmable delays applied using three VIRTEX-II XC2V40 Field Programmable Gate Arrays (FPGAs) [5] called Delay FPGAs. A data-spy system is also implemented in the Delay FPGAs with the purpose of monitoring the raw data from the tracker before processing.

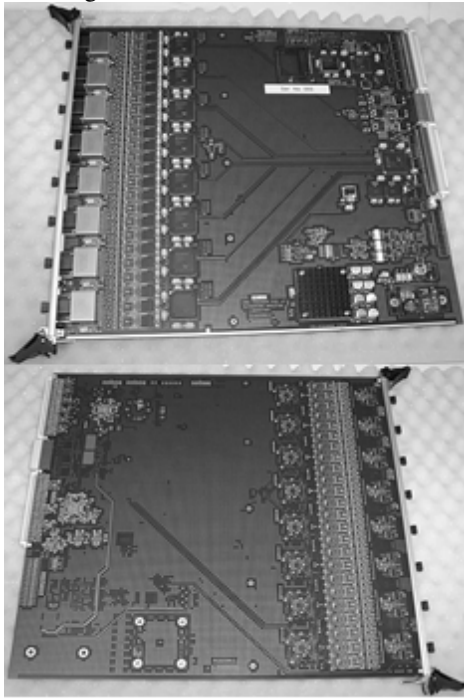

Fig. 2: The CMS Tracker FED card primary side (top) and secondary side (bottom). Eight front-end units each consisting of an Opto-Rx module serving 12 fibres corresponding to signals from 24 APV25s, 6 dual 10 bit AD9218 ADCs, 3 VIRTEX-II XC2V40 FPGAs and one VIRTEX-II XC2V1500 FPGA are shown in the front section of the card. The VME 64x interface is located adjacent to the VME J1 connector. The data are collected in the back-end FPGA (VIRTEX-II XC2V2000) and send to the DAQ system via the S-LINK64 interface through the VME J2 connector.

The digitized data are then sent via the delay FPGAs to one VIRTEX-II XC2V1500 FPGA [5], called the Front-End FPGA for processing. In the Front-End FPGA the data are synchronized and only data coming within one predefined time window is accepted. The data are reordered and programmable pedestals are subtracted from the data. The common mode offset is computed by finding the median charge deposition of the strips corresponding to one APV25 and is subtracted from the data. The corrected data are then processed by a cluster finding algorithm which searches for clusters of neighboring strips where the signal over noise is larger than a threshold or isolated strips with signal over noise larger than a second higher threshold. Both thresholds are programmable via VME. The cluster data are stored in a 2kB FIFO, in the FPGA, compressed to eight bits since a typical minimum ionizing particle will register about 80 counts. This process reduces a total FED input raw data rate of 3.4 GB/sec to roughly 50 MB/sec per %-strip-occupancy. The average strip occupancy rates in different regions of the tracker will vary between ~0.5% and 3% in high luminosity LHC running.

### B. The FED Back-End FPGA

The data from the eight Front-End FPGAs are transmitted via 4-bit point-to-point connections running at 160 MHz to a VIRTEX-II XC2V2000 FPGA [5], called the Back-End FPGA, seen in Fig. 2 near the J2 VME connector. A TTC-Rx chip receives the Level-1 Trigger and the 40 MHz clock, coming from the TTC system [6], via a fibre from the FED front panel and passes them to the Back-End FPGA. Using this information the Back-End FPGA constructs an event header that includes the Level-1 trigger information, bunch crossing information and the CRC code to check for data corruption. The FED event record is then constructed by appending the data from the Front-End units to the header. The FED can be programmed via VME to transmit either cluster data or simply the raw data for debugging purposes albeit at a lower trigger rate. The events are stored in an external 2 MBytes deep buffer consisting of a pair of Quad Data Rate (QDR) SRAMs to account for trigger rate fluctuations before they are routed to the VME J2 connector on the way to the CMS DAQ. The QDRs are also accessible via VME64x that provides a second but slower readout path. The Back-End FPGA is also responsible for transmitting via the J2 VME connector a 4-bit word for handshaking with the CMS Trigger Control System (TCS) [7]. Depending on the FED buffers status this feedback information is interpreted by the TCS as READY (to take more triggers), BUSY (stop triggers), WARNING-OVERFLOW (reduce triggers), OUT-OF-SYNCH, ERROR (Reset). The Trigger system then responds by sending the FED the appropriate information via the TTC-Rx path. It is intended that future firmware upgrades will also include monitoring firmware in the Back-End FPGA design. Such firmware will monitor quantities such as occupancy, number of channels out of synchronization and buffer-sizes. This information will be routed via VME to online monitoring Software.

### C. The FED Interface to the CMS DAQ

Each FED transmits its complete event records to the CMS DAQ via the CMS Front-end Readout Links (FRL) [8] which implement the 400 Mbyte/sec S-LINK64 protocol [9]. To interface with the FRL system a 6U VME transition card, shown in Fig. 3, has been designed. The card is designed to be plugged in the backside of the VME crate with access the J2 VME connector. The transition card receives the event data record and the 4 TCS bits in single ended format and an 80 MHz differential clock. The S-LINK64 transmitter mezzanine card [8] is mounted on the top of the transition card and receives the FED data. The transition card also receives the 4 TCS bits, converts them to LVDS and transmits them via the RJ45 connector to be combined with those of the other FEDS via the FMM [10] cards on the way to the APV-Emulator cards, described in the next section, which form the Tracker Interface to TCS.

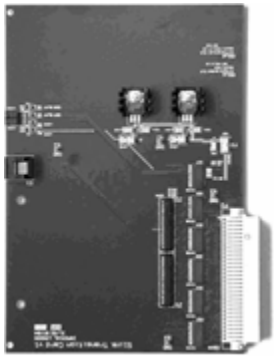

Fig. 3: The FED S-LINK64 transition 6U VME card. The S-LINK64 transmitter mezzanine is plugged on the black connector shown next to the VME buffers. The RJ45 connector transmits the TCS signals.

### III. THE TRACKER READOUT INTERFACE TO THE CMS TRIGGER SYSTEM

The CMS tracker readout is designed to operate up to a maximum trigger rate of 100 kHz. To accommodate for Poisson fluctuations of the trigger rate the APV25 has a 31-cell deep FIFO, which stores the pipeline addresses waiting to be read out. In peak mode this allows the analogue information from 31 triggers to be stored in the pipeline. De-convolution mode, used in normal CMS physics running, requires 3 samples per trigger and therefore the analogue information for a maximum of 10 triggers can be stored. If this 10-event buffer overflows the APV25 need to be reset. This can only be done globally for the entire tracker leading to a considerable loss of data. Hence, there is a need to transmit information about the status of the APV25 front-end buffers to the trigger system which could then act to prevent the APV25 buffers from overflowing by adjusting the trigger rate. Given the distance between the Silicon Tracker and the CMS counting room it is not possible to transmit this information from the actual APV25s on the detector in time. This task is accomplished by the APV Emulator (APVE) board shown in Fig. 4. The APVE design utilizes the property of the APV25 architecture where the buffer occupancy depends only on the FLT rate and emulates the status of the APV25 buffers. It can perform this task both by using an actual on-board APV25 and by using a VHDL version of the APV25 logic running on a VIRTEX II XC2V1000 FPGA. Additional logic implemented in VHDL on the XC2V1000 FPGA monitors the occupancy the APV25 buffer by counting the number of the first level triggers received and comparing it with the number of APV25-events coming out. If the APV25 Buffers are about to become full the APVE will assert a WARNING-OVERFLOW signal to the GT and when that buffers are full it will assert a BUSY. The time that the WARNING-OVERFLOW signal is asserted is programmable. The total feedback time is about 75 ns. To perform this task the APVE has an interface to the CMS TCS system and provides also the READY, ERROR and OUT-OF-SYNCH signals. In addition to its main task the APVE also performs two more tasks. It receives the global logical OR of the TCS signals sent by the FEDs via the FMM cards and combines them with those from the APV25 logic to form a global summary for the entire tracker system and transmits it to the TCS. The third task of the APVE is to check overall the synchronization the CMS tracker system. This is done by transmitting the pipeline address that comes from its emulation circuitry and corresponds to a given FLT decision to the FEDs via the CMS TTC system. The FEDs compare the pipeline address from the APVE with that coming from the header of the APV25 data on the detector and test if the data from the given section of the tracker is in synch with the Trigger system. The APVE will be installed in the same rack as the Global First Level Trigger (GT) and receives the trigger information via a dedicated connection to the GT.

### IV. THE FED TESTER CARD FOR TESTING THE FED

Testing the FEDs requires a device that emulates both the tracker analogue optical signals and the trigger digital signals required for the proper operation of the FED. This is a challenging task because it requires that the CMS Trigger and DAQ environment is created 'on the bench' providing signals whose shape and timing is identical to those that will eventually be available at CMS. Such a device is the FED Tester Card (FTC) which has been designed and manufactured and shown in Fig. 5.

#### A. The FED Tester Architecture

The FTC uses two XC2V1000 FPGAs for its internal logic. The XC2V1000 RAMs are used to store the test patterns which are then clocked out at 40 MHz. The phases of the internal clocks are controlled using the XC2V1000 digital clock managers and are used to control the timing of the test patterns. The test-pattern data are converted to analogue using three AD9753 DACs [4] driving three programmable AD8108 [4] analogue cross-point switches which act as three-to-eight fan-outs. Each DAC is driving any of the three cross point switches. The outputs of the switches drive 8 CMS tracker Opto-Tx modules which emulate the analogue optical data for 24 FED channels. Four FTCs are required to fully test all 96 FED input channels. Hence, for synchronization reasons each FTC can be programmed to act either as a clock and trigger master, driving the other three FTCs, or as a slave which will expect trigger and clock from a master FTC. The FTC design is implemented on a VME 9U 400mm card which acts as an A24D16 VME slave.

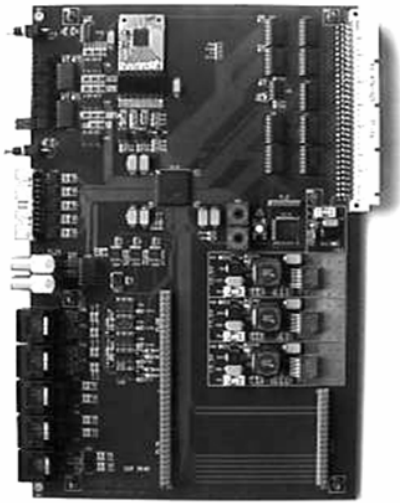

Fig. 4: The 6U VME APV25 Emulator card. The APV25 ASIC can be seen mounted on a copper PCB at the top. The FMM inputs and TCS outputs are seen at the lower left side.

### B. Temperature control of Opto-Tx modules

Testing the FED inputs with a range of signals extending to values that could be smaller than the equivalent deposition of a minimum ionizing particle is a demanding task because it requires that the temperature of the Opto-Tx modules is held stable to better than 0.5 degrees. This is achieved by keeping the Opto-Tx at a temperature higher than room temperature, by heating it with a resistor, and a digital feedback loop, employing a PID algorithm, which measures the temperature and controls the current through the resistor.

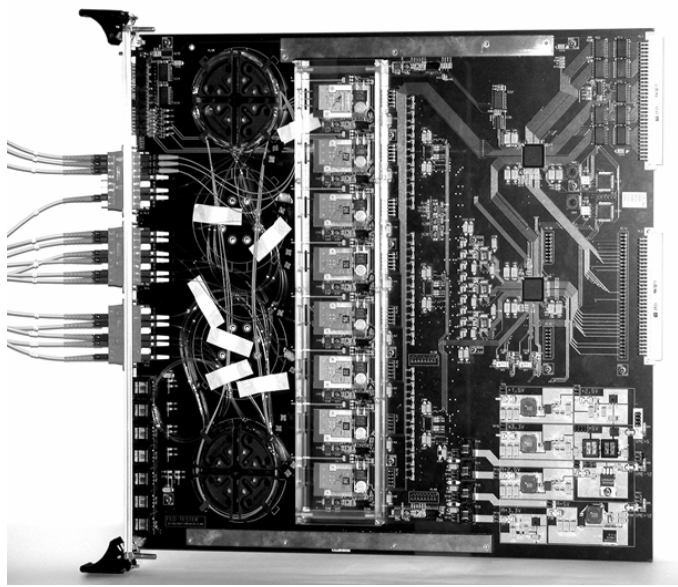

Fig. 5: The 9U VME Fed Tester Card. Each card serves 24 Fibre connectors shown in the Front Panel. The Card interfaces with a PC via the J1 VME interface. The J2 VME connector user defined pins are used to transmit the clock the FEDs. The RJ45 connectors in the front panel can be used to connect and synchronize the card with other Fed Tester Cards in the same VME crate.

## V. FED Performance Tests And Discussion

Shown in Fig. 6 are measurements of peak RMS noise at the beginning of the copper analogue section with and without the Opto-Rx receiver mounted on the FED PCB. The overall noise level is less than an ADC count. The Opto-Rx contribution to noise is clearly seen in the first 24 channels which had 2 Opto-Rx modules mounted.

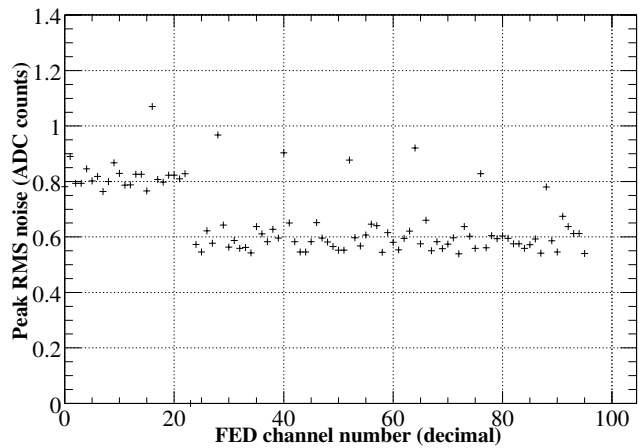

Fig. 6: Peak rms noise versus FED channel. For this test only the first 24 channels are connected to Opto-Rx module outputs.

The FTC has been used to inject data patterns at the FED inputs. The data were read via VME and detailed tests of the FED zero suppression algorithms have been preformed. The tests have established the full functionality of the Front-End FPGA firmware.

Data patterns generated under software control in the Back-End FPGA and from FTC have been used to study the FED output data transfer rate as a function of the cluster data event size for a trigger rate of 100 KHz. Seen in Fig. 7 is the size of the FED buffer as a function of the output data rate. The buffer size increases until the output rate saturates at 469 Mbytes/sec due to the S-LINK64 asserting the backpressure bit indicating that the link had reached its throughput limit. This is well above the design specifications for the link which is designed for a 200 MByte/sec average and 400 Mbytes/sec peak rates.

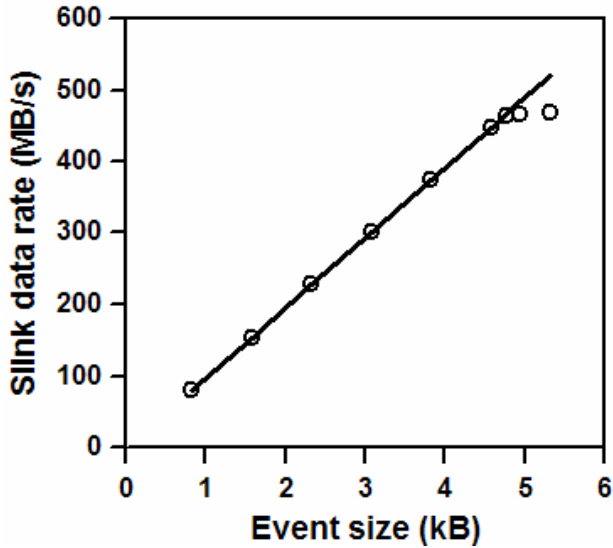

Figure 7: SLINK-64 data transfer rate versus the cluster data event size.

A high tracker occupancy during pp-running will result into an increased FED buffer occupancy which may result in a WARNING-OVERFLOW or a BUSY transmitted to the Trigger system. This will induce an undesired readout dead-time. Hence, the conditions under which this occurs have been studied. Using the FTC, test data has been injected at the FED inputs emulating different values of the tracker occupancy. The trigger rate was Poisson distributed with an average of 100 KHz. Due to the fact that the final buffer monitoring firmware was not yet implemented, a direct measurement of the dead time was not possible and the FED buffer status was determined by reading the FED status register bit which indicated that a buffer overflow had indeed occurred. The results of this study are shown in Fig. 8 where the tracker occupancy is plotted versus the average, over 5000 tests, number of events taken before an overflow had occurred. The tracker occupancy expected at LHC for high luminosity running ($10^{34} cm^{-2} sec^{-1}$) is below 3% per strip. As seen in Fig. 8 the FED will induce dead time for occupancy larger then 6.25% which is a more than factor of two larger than the maximum occupancy expected. Based on the results shown, the CMS tracker FEDs are performing according to the design specifications.

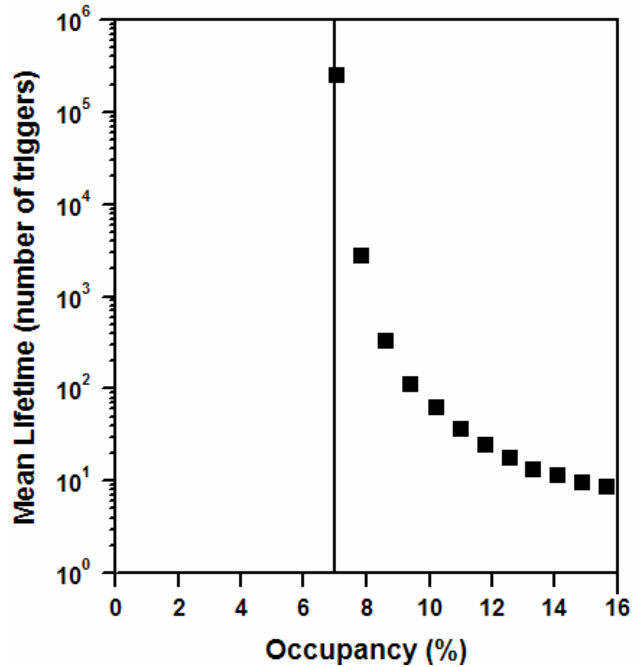

Figure 8: The average number of events taken before a FED buffer overflow occurred versus the tracker occupancy.

## VI. ACKNOWLEDGMENTS

The authors wish to thank C. Schwick and D. Gigi from the CMS DAQ group for providing information help and support for the S-LINK64 hardware. The authors would also like to thank PPARC for the continued financial support.

## VII. REFERENCES

[1] M. Raymond et. al., *The CMS tracker APV25 0.25 μm CMOS Readout chip,* Sixth Workshop on Electronics for LHC Experiments, CERN/LHCC/2000-041, p130.
[2] J. Troska et al., *Optical readout and control systems for the CMS tracker,* IEEE Transactions on Nuclear Science, Volume: 50, Issue 4, Aug. 2003, 1067-1072.
[3] *Optical Links for CMS,* cms-tk-opto.web.cern.ch/cms-tk-opto/ .
[4] Analogue Devices, Inc., *www.analog.com*.
[5] Xilinx, "Virtex-II Platform FPGA Handbook", *www.xilinx.com* .
[6] *Timing, Trigger and Control Systems for the LHC,* ttc.web.cern.ch/TTC/intro.html.
[7] J. Varela, *CMS L1 Trigger Control System*, CMS Note, 2002-033.
[8] The Trigger and Data Acquisition Project, Volume II. Data Acquisition & High-Level Trigger", CERN/LHCC 2002-26, CMS TDR 6.2, December 15, 2002.
[9] H.C. van der Bij et al., *S-LINK, a data link interface specification for the LHC era,* IEEE Transactions on Nuclear Science, Volume: 44 Issue: 3 , June 1997, 398-402.
[10] A. Racz et al. *The final prototype of the FMM for readout status processing in CMS DAQ,* Proceedings of the LECC2004, Boston 2004.